\documentstyle[11pt]{article}

\input epsf

\def\pa{\partial}
 
\def\g{\gamma} \def\G{\Gamma}
\def\a{\alpha} 
\def\b{\beta} 
\def\d{\delta}

\def\e{\epsilon}

\def\l{\lambda} \def\L{\Lambda}
\def\m{\mu} 
 
\def\x{\chi} 
\def\p{\pi}

\def\mn{{\mu\nu}}

\def\ul{\underline}
\def\be{\begin{equation}}
\def\ee{\end{equation}}

\begin{document}


\begin{flushright}
hep-th/9602069\\ \vspace{-.2in}
BRX TH--390
\end{flushright}

\vspace{-.4in}

\begin{center}
{\Large\bf The Gauged Vector Model in Four-Dimensions:\\
Resolution of an Old Problem?}

David L. Olmsted\footnote{email address:
olmsted,schnitzer@binah.cc.brandeis.edu}\\
and \\
Howard J. Schnitzer$^{1,}$\footnote{Research supported in 
part by the DOE under grant DE--FG02--92ER40706.}\\
Martin Fisher School of Physics\\
Brandeis University, Waltham, MA 02254
\end{center}

\begin{quotation}
{\bf Abstract}: A calculation of the renormalization group
improved effective potential
 for the gauged U(N) vector model,
coupled to $N_f$ fermions in the fundamental representation, 
computed to leading order in
1/N, all orders in the scalar self-coupling $\l$, and lowest 
order in 
gauge coupling $g^2$, with $N_f$ of order $N$, is presented. 
It is shown that the theory has two phases, one of which is  
asymptotically free, and the other not, where the 
asymptotically 
free phase occurs if 
$0 < \l /g^2 < \frac{4}{3} (\frac{N_f}{N} - 1)$, and 
$\frac{N_f}{N} < \frac{11}{2}$.  In the 
asymptotically 
free phase, the effective potential behaves qualitatively 
like the tree-level potential.  In the other phase, the 
theory exhibits all the 
difficulties of the ungauged $(g^2 = 0)$ vector model.  
Therefore the theory appears to be consistent (only) in 
the asymptotically free phase.
\end{quotation}

\newpage

\noindent{\bf 1. ~Introduction}

The search for non-perturbative methods in quantum field 
theory remains a central issue of the subject.  Although 
great progress has been
made recently using duality \cite{001}, there is still 
considerable interest in other approaches to strong-coupling 
questions, particularly as
the new methods are limited to supersymmetric theories at 
present.  One of the other techniques most frequently 
considered is the 1/N expansion for 
a theory with internal symmetry, continued to O(N) or U(N) 
for example. Applications include 't Hooft's analysis of 
gauge theories \cite{0022}, the 
demonstration of string behavior in two-dimensional QCD 
\cite{0023}, and of 
O(N) invariant $\l\phi^4$ theory \cite{003,004}.

The 1/N expansion for $\l\phi^4$ theory (in 3+1 dimensions) with
O(N) symmetry (the so-called vector model) has been extensively
studied as a renormalized field theory \cite{003,004}.
However, the renormalized vector model encounters a number 
of problems \cite{004}.  Among these are:\\
(1)~ The effective potential of the theory is double-valued,
where the lower branch of the potential exhibits \ul{unbroken} 
internal symmetry at it's minimum, {\it i.e.}, $\langle \phi_a 
\rangle = 0$  [See Fig. 3 of Abbott, {\it et al.}, Ref. 5.]  
This 
phase is tachyon-free in all orders of the 1/N 
expansion. The upper branch of the effective potential does 
allow a spontaneous broken symmetry, but at the expense of 
the appearance of 
tachyons, which signals a decay to the lower energy phase.  
In higher 
orders of the 1/N expansion, the upper branch of the effective 
potential becomes everywhere complex.\\
(2)~ The effective potential has no lowest energy bound as the 
external field $\phi \rightarrow \infty$.  The tachyon-free 
phase ({\it i.e.}, with $\langle \phi_a \rangle = 0$) tunnels non-perturbatively to this
unstable vacuum.

The primary motivation of this paper is to provide a plausible 
resolution of the problems encountered by Abbott, {\it et al.} 
[5], although there may 
be other solutions as well.  It is probably relevant that 
the ungauged $\l\phi^4$ theory in four dimensions seems to be 
trivial \cite{007}, but 
this feature is not the focus of this paper.

These difficulties make the renormalized vector model, evaluated
in the 1/N expansion, unsuitable for phenomenology.  There was 
also interest in studying this model in the double-scaling 
limit \cite{005}, where one 
considers the correlated limit, $N \rightarrow \infty$ and 
$\l \rightarrow \l_c$, where $\l_c$ is a critical value of the 
coupling.  Unfortunately, just at the critical point, the 
effective potential becomes
everywhere complex \cite{007}, so that particular application 
of the vector model is also not possible.

One response to these problems is to consider a cutoff version 
of the vector model in the 1/N expansion \cite{008,009}.  In 
that case a viable phenomenology, with spontaneously broken 
symmetry and no tachyons, does exist to leading order in 1/N.  [Unfortunately, a double-scaling
limit is still not possible even in the cutoff-version of 
the vector model \cite{009,010}.]  Since a cutoff mass $\L$ 
represents an energy scale above which the scalar fields 
have significant interactions with other
degrees of freedom, the cutoff $\l\phi^4$ theory cannot 
be regarded as a closed system, in that there are degrees 
of freedom which have been neglected,
and in some sense incorporated into the cutoff $\L$.

The question is then how should one couple the scalar 
fields to additional degrees of freedom, so that the 
system is consistent with just these degrees 
of freedom and \ul{no} cutoff?  In this paper, we argue 
that one way this can be accomplished in the 1/N expansion 
is by gauging the scalar fields,
and adding $N_f$ massless fermions in the fundamental 
representation, if the scalar self-coupling satisfies 
$0 < \l < \frac{4}{3} (\frac{N_f}{N} -1)g^2$ and 
$N_f/N < 11/2$ in the large 
$N$ limit.  If not, one returns to all the difficulties 
of the ungauged model.  
Notice that the model is {\bf not} asymptotically free 
for $N_f =0$ in the large $N$ limit.

The gauged vector model in the 1/N expansion was previously 
considered by Kang \cite{011}.  
However, his calculation is inadequate for our purposes, 
as the renormalization scheme chosen would not allow for a 
conventional Higgs mechanism.  More importantly, the 
conclusions drawn by Kang were not 
reliable \cite{012}, as they depended on features of the 
effective potential outside the domain of validity of the 
calculation, as evidenced by large logarithms.  In our 
work we remedy both of these difficulties.

In Sec. 2 we formulate the gauged vector model coupled 
to fermions 
in the fundamental representation, and solve for the
effective potential to leading orders in 1/N, to all 
orders in $\l$, 
and to leading order in $g^2$.  A renormalization group 
improved effective potential is constructed and analyzed 
in Sec. 3.  It is argued that the theory has two phases.  
If $\l /g^2$ is small enough, the model is 
asymptotically free and the theory is consistent, in that 
the difficulties found by Abbott, {\it et al.}, \cite{004} 
are absent in this phase.   In Sec. 4 there is a brief 
discussion of the issue of gauge invariance.

\vspace{.2in}

\noindent{\bf 2.~ The Gauged Vector Model}

\renewcommand{\theequation}{2.\arabic{equation}}

Let us consider a renormalizable theory of gauged complex 
scalar fields in the 
fundamental representation of U(N), with gauged-fixed 
Lagrangian density \cite{011}, and $N_f$ massless fermions 
in the fundamental representation as well.
\begin{eqnarray}
N^{-1}{\cal L} & = & |\pa_\m \phi + igA_\m \; \phi |^2 
\nonumber
\\
& + & \frac{1}{2\l} \: \x^2 - \frac{\m^2}{\l} \: \x - \x \; |
\phi |^2 \nonumber\\
& - & \frac{1}{4} \: Tr (F_\mn F^\mn ) - \frac{1}{2\xi} \: Tr
(\pa_\m A^\m )^2 \nonumber \\
& - & Tr \left\{ \pa_\m C^* (\pa^\m C + ig [A^\m,C] \right\}
\nonumber \\
& + & i \, \sum^{N_f}_{i=1} (\bar{\psi}_i \g \cdot D \psi_i )
\end{eqnarray}
In (2.1) $\phi$ and $\psi_i$ transform in the fundamental 
representation of
U(N), the gauge field 
(ghost) $A_\m \; (C)$ transform in the adjoint, $\x$ is a
U(N) singlet, and $D$ is the covariant derivative. 
The field $\x$ serves as a Lagrange parameter,
which if eliminated, 
reproduces the usual $\l\phi^4$ scalar self-interactions.  
[The coupling constants and fields have been rescaled so 
that N is an overall factor of the Lagrangian, and hence 
1/N is a suitable expansion
parameter.]  Note that there is no Yukawa coupling between
$\phi$ and $\psi$, since both are in the fundamental
representation. In the absence of the gauge couplings, (2.1) 
reduces to the usual vector 
model with U(N) symmetry \cite{003,004}, together with
$N_f$ free fermions.

In this section we present the results of a calculation 
of the effective 
potential derived from (2.1) to leading order in 1/N, to 
all orders in $\l$, 
and leading order in $g^2$.   It is convenient to work in  
Landau gauge ($\xi = 0$), so that the gauge parameter will 
not be renormalized.  The resulting effective potential 
is renormalized using modified minimal subtraction, 
with the relevant relations 
between bare and  renormalized quantities to the order 
indicated above being
\begin{eqnarray}
d & = & 4-2\e \nonumber \\
M^\e \phi_b & = & Z^{1/2}_\phi \, \phi_r \nonumber \\
M^\e \psi_b & = & Z^{1/2}_\psi \, \psi_r \nonumber \\
\x_b & = & Z^{-1}_\phi \, \x_r \nonumber \\
M^{-\e}\,g_b & = & g_r \nonumber \\
M^{2\e} \left(\frac{\m^2}{\l} \right)_b & = & Z_\phi 
\left(\frac{\m^2}{\l} \right)_r \nonumber \\[.05in]
M^{2\e} \left(\frac{1}{\l} \right)_b & = & Z^2_\phi 
\left( \frac{1}{\l} \right)_r \; - \; \frac{1}{16\pi^2\e} 
\;
 - \; \frac{1}{16\pi^2} \left( \frac{g^2}{16\pi^2} \right)\;
\left(\frac{3}{\e^2} + \frac{4}{\e} \right) 
\nonumber \\[.05in]
Z_\phi & = & 1 \; + \; \frac{g^2_r}{16\pi^2} \; 
\left( \frac{3}{\e} \right)\; .
\end{eqnarray}
The subscripts $b$ and $r$ refer to bare and renormalized
quantities respectively, while $M$ is an arbitrary 
renormalization mass-scale. [Note that in the 1/N 
expansion it is natural to renormalize 1/$\l$ rather 
than $\l$.]  The resulting renormalized effective potential 
is
\begin{eqnarray}
N^{-1}V & = & \frac{\m^2}{\l} \: \x \; + \; \x \phi^2 \; - \;
\frac{1}{2\l} \; \x^2 \nonumber \\
& + & \frac{1}{16\pi^2} \; \x^2 \; \left[\frac{1}{2} \: \ln \,
\left(\frac{\x}{M^2} \right) \; - \; \frac{3}{4} \right]
\nonumber \\[.05in]
& - & \frac{1}{16\pi^2} \, \left(\frac{g^2}{16\pi^2} \right) \,
\x^2 \, 
\left[ \frac{3}{2} \, \ln^2 \, \left( \frac{\x}{M^2} \right) \; -
\;
7 \, \ln \, \left( \frac{\x}{M^2} \right) \; + \; c \: \right]
\end{eqnarray}
where $c$ is a numerical constant not relevant to our order. 
[The subscripts $r$ will be omitted in all that follows.  
Note that the fermions do not contribute to (2.3) to leading
order in $N$ and $g^2$, due to the absence of a Yukawa coupling.  
For convenience we write $\phi^2$ instead of $|\phi|^2$.]  One 
is interested in the ultraviolet behavior of the effective 
potential to see if the
difficulties found by Abbott {\it et al.}, \cite{004} have been
eliminated. 
 However, when $\x /M^2 >> 1$, one encounters 
large logarithms which make (2.3) unreliable in that region. 
Therefore we 
consider the renormalization group improved effective potential,
which will 
provide an effective potential which is independent of $M$, to
the order we 
are working, and suppress the dependence on large logarithms.

To this order, we want an effective potential which satisfies
\begin{eqnarray}
0 & = & M \; \frac{dV}{dM} \nonumber \\
&  = & \left[ M \, \frac{\pa}{\pa M} \; + \; \b_{1/\l} \;
\frac{\pa}{\pa 
(1/\l)} \; + \; \b_g \: \frac{\pa}{\pa g} \right. \nonumber
\\[.05in]
& + & \left. \b_{\m^2/\l} \; \frac{\pa}{\pa (\m^2 /\l)} \; - 
\:\g_\phi \: \phi^2 \: \frac{\pa}{\pa \phi^2} \; + \; \g_\phi \;
\x \: \frac{\pa}{\pa \x} \right] V \; ,
\end{eqnarray}
and agrees with (2.3) when expanded in $g^2$ and $\ln (\x /M^2)$. 
Equation (2.4) does not depend on $\psi$, as there are no external 
fermion insertions.  The 
$\b$-functions and anomalous dimensions obtained from (2.2) are, 
to leading order in $N$
$$
\b_{1/\l} \; = \;  M \; \frac{d}{dM} \; \left(
\frac{1}{\l}\right) 
\; = \; \frac{1}{16\pi^2} \; \left[
\frac{32\p^2\e}{\l} \; + \; \frac{12g^2}{\l} \; - 2 \; - \;
\frac{g^2}{\p^2}
\right]
\eqno{(2.5{\rm a})}
$$
$$
\b_g \; = \; M \; \frac{d}{dM} \; g \; = \; -\e g \; - 
g^3/16\pi^2 \left( \frac{22}{3} - \frac{4}{3} \:
\frac{N_f}{N} \right) + {\cal O} (g^5)
\eqno{(2.5{\rm b})}
$$
$$
\b_{(\m^2/\l )} \; = \; M \; \frac{d}{dM} \; \left(
\frac{\m^2}{\l} \right) \; 
=\; (2\e - \g_\phi ) \:
\left( \frac{\m^2}{\l} \right)
\eqno{(2.5{\rm c})}
$$
$$
\g_\phi \; =\; M \; \frac{d}{dM} \;\ln Z_\phi
\; =\; \frac{-6g^2}{16\p^2} \; + \; {\cal O}(g^4) \; .
\eqno{(2.5{\rm d})}
$$
We therefore consider $N_f$ of order $N$. 
It is useful to define
$$
\g^\prime \; = \; \frac{\g_\phi}{(1-\g_\phi /2)}
$$
so that $\x^2\; (\frac{\x}{M^2})^{\g{^\prime}}$ is a
renormalization group 
invariant.  Note that the conventional $\b_\l$ function is
related to (2.5a) 
by
$$
\b_\l \; =\; -\l^2 \, \b_{1/\l}
$$
so that
\renewcommand{\theequation}{2.\arabic{equation}}
\setcounter{equation}{5}
\be
16\pi^2 \b_\l = -32\p^2\e\l \; + \; 2\l\,
\left( \l - 6g^2 + \frac{g^2\l}{2\pi^2} \right) \; .
\ee

Let $M_0$ be the mass-scale at
which the 
coupling constants and ``composite" field $\x$ are defined, and
\begin{eqnarray}
\l_0 & = & \l (M_0) \nonumber \\
\x_0 & = & \x (M_0; \phi^2)
\end{eqnarray}

\noindent{\bf 3. Renormalization Group Improved Effective 
Potential}

\noindent{\bf A. Effective Potential}

Let us consider the renormalization group improvement of 
the effective potential (2.3)

\renewcommand{\theequation}{3.\arabic{equation}}
\setcounter{equation}{0}

From (2.5b)
\be
g^2 (M) = g^2_0 
\left[ 1 + \frac{4}{3} \left( 11-2 \; 
\frac{N_f}{N} \right)
 \: \frac{g^2_0}{16\pi^2} \: \ln 
\left( \frac{M}{M_0} \right)\right]^{-1}
\ee
where
\be
g_0 = g(M_0) \; .
\ee
From (2.5a) and (2.5b) we have
\be
M \: \frac{d}{dM} \left[ \frac{16\pi^2}{\l} -
\left( \frac{2}{12-A} \right) \frac{16\pi^2}{g^2} \right]
=
\frac{12g^2}{16\pi^2} 
\left[ \frac{16\pi^2}{\l} - \left( \frac{2}{12-A} \right)
\; \frac{16\pi^2}{g^2} \right] + {\cal O} (g^2)
\ee
where
\be
A = \frac{4}{3} \left( 11 -2\: \frac{N_f}{N} \right)
\ee
We see that there is a phase-boundary when
\be
\l = \frac{4}{3} \: \left( \frac{N_f}{N} - 1 \right)
g^2
\ee

A graph of $\l$ versus $g^2$ is shown in Figure 1, with the 
renormalization group flows indicated on the graph.  Note the
two-phase structure of the theory, with the line $\l = 4/3 \:
(\frac{N_f}{N} - 1) g^2 = C\: g^2$ in the $(\l , g^2)$ plane 
separating the two-phases.  For $\l < C \: 
g^2$, the theory is asymptotically free, while if 
$\l > C \: g^2$, the 
theory is not, since $\l \rightarrow \infty$ in the ultraviolet 
in this phase even though $g^2 \rightarrow 0$. Thus the 
qualitative ultraviolet behavior of the theory in the 
phase $\l > C \: g^2$  is similar to that of the ungauged 
theory.  If the initial conditions for the renormalization 
group are chosen such that $\l_0 = C \: g^2_0$, then 
$\l/g^2 = C$ throughout the renormalization group flow, 
to the order we are working.  
Note that if $g^2_0 / 16 \pi^2 \ll 1$, then also 
$\l_0/16\pi^2 \ll 1$ in 
the asymptotically free phase; which {\it \`{a} posteriori} 
is in the domain of perturbation theory.

The solution to (2.4) and (2.5), with (2.3) as boundary 
conditions, gives the renormalization group improved
effective potential\footnote{The condition for broken 
symmetry, $\m^2 /\l < 0$ with $\frac{\pa V}{\pa \phi^2} = 0$ 
requires $\chi = 0$ in both the 
tree level potential, and in (3.5).  Also, the massless 
fermions do not contribute to the vacuum energy to leading 
order in $N$ and $g^2$.  Therefore, the vacuum energy
is zero and does not need separate renormalization group 
improvement. See 
\cite{019}. We thank B. Kastening for 
raising this point.} 
\begin{eqnarray}
N^{-1} V & = & \frac{\m^2}{\l} \: \chi + \chi \phi^2 
\nonumber \\[.1in]
& - & \frac{1}{2} \: \chi^2 
\left\{ \frac{1}{\l} \: - \: 
\left( \frac{2}{12-A} \right)
\frac{1}{g^2}
\left[ 1 - \left( 1 + \frac{A}{2} \: \frac{g^2}{16\pi^2} \ln 
\left( \frac{\chi}{M^2}\right) \right)^{\textstyle{\frac{A-12}{A}}} 
\right]\right\} 
\end{eqnarray}
where $A$ is given by (3.4).
In solving the renormalization group equation (2.4) one 
matches only the leading logarithms of (2.3), since one 
has no control of the sub-leading 
logarithms, to the order we are working.  Thus 
$M \frac{dV}{dM}$ is not identically zero for (3.6), 
but is zero to the order of accuracy required 
of our approximations.  The gap equation $\pa V/\pa 
\chi =0$ means that
\begin{eqnarray}
\phi^2 = \frac{-\m^2}{\l} +  \chi
\left\{ {\rule{0mm}{7mm}}\frac{1}{\l} \right. & - & 
\left( \frac{2}{12-A} \right) 
\frac{1}{g^2}
\left[ 1 - \left( 1 + \frac{A}{2} \: \frac{g^2}{16\pi^2} \ln 
\left( \frac{\chi}{M^2} \right) 
\right)^{\textstyle{\frac{A-12}{A}}} 
\right]  \nonumber \\
& - & \frac{1}{32\pi^2} \,
 \left[ 1 + \, \frac{A}{2} \left. \frac{g^2}{16\pi^2} \ln 
\left( \frac{\chi}{M^2} \right) 
\right]^{\textstyle{\frac{-12}{A}}}
\right\} 
\end{eqnarray}
Thus, inserting (3.7) 
into (3.6)
\be
N^{-1} V = \frac{1}{2} \: \chi^2 
\left\{ \frac{1}{\l} \: - \: 
\left( \frac{2}{12-A} \right)
\frac{1}{g^2}
\left[ 1 - \left( 1 + \frac{A}{2} \: \frac{g^2}{16\pi^2} 
\ln \left( \frac{\chi}{M^2} \right) 
\right)^{\textstyle{\frac{A-12}{A}}} 
\right]\right\}  \; ,
\ee
where we dropped subleading terms in $g^2$ $\ln (\chi /M^2)$, 
which can be neglected to the order we are working. The 
coefficient of the subleading term in (3.7) is not determined 
by our computation, because it corresponds to a higher order 
term in (3.6).  However, since what we really have is (3.6) and 
$\pa \chi /\pa V = 0$, (3.7) must be used as shown in numerical
computations. One must resort to a numerical evaluation for 
$V(\phi^2)$ since (3.6) with (3.7) cannot be evaluated 
analytically.

Note from Figure 1 that the model has two phases, with 
the phase boundary given by\footnote{Note that there is no asymptotically free phase for $N_f =0$.  A factor of two 
error in (3.1) led to the opposite conclusion in an
earlier version of this paper.}
\be
0 < \l = \frac{4}{3} \left( \frac{N_f}{N} - 1\right)g^2
\ee
which depends on $1 < ( \frac{N_f}{N} ) < \frac{11}{2}$,
where the upper-bound is required so as to maintain asymptotic
freedom, as can be seen from (3.1).  Since $g^2 (M)$ has a
Landau singularity in the infrared region, a Landau pole 
appears in the infrared region of the effective potential.
The infrared pole should be regarded as a signal of the
confinement of colored degrees of freedom, and not a
fundamental flaw in the model.  In the asymptotically free
phase, where $\l < \frac{4}{3} (\frac{N_f}{N} - 1)g^2$,
the effective potential has a lowest energy bound in the
ultraviolet, while if $\l > \frac{4}{3}(\frac{N_f}{N} - 1)g^2$
there is no lowest energy bound in the ultraviolet.
[With the presence of the Landau pole, one cannot discuss
a lowest energy bound in the infrared in a meaningful way.]
We present in Figs. 2 and 3, the effective potential
$V(\phi^2)$ for the two phases, for $\m^2 < 0$.  The
infrared singularity occurs in a range of $\phi^2$ many
orders of magnitude smaller than the scale of the
figures, so it disappears in the interval between two of 
the numerically computed points.  The figures
emphasize the possibility of spontaneous symmetry breaking,
and stability or lack of stability of the phases.  [In
Figs. 2 and 3, $g^2 (M_0)/16\pi^2$ is chosen sufficiently
small so that the Landau pole is in the extreme infrared
region.]

The phase boundary is given by (3.8), with corrections of
${\cal O} (g^4)$ expected.  For $g^2 (M_0)/16\pi^2$ sufficiently
small, these corrections are not expected to shift the
phase-boundary in any significant way.  We observe from
(3.3) and (3.4) that the {\bf ratio} $\l /g^2$ does not
run at the phase boundary, although both $g^2 (M^2)$ and 
$\l (g^2)$ flow to zero in the ultraviolet.

\noindent{\bf B. ~Vector Meson Spectrum and Confinement}

As we have just discussed, Figs. 2 and 3 give a description 
of the effective potential for our model for $\m^2 < 0$, 
in the two phases of the theory.  The results for 
$\m^2 >0$ are qualitatively 
similar, except the $\phi^2 = 0$ axis in Figs. 2 and 3 
is shifted to the right, so that no spontaneous symmetry 
breaking takes place. The Landau singularity, which is not
shown in the figures, signals that the theory likely will 
confine.

It is claimed that in non-abelian gauge theories with matter in 
the fundamental representation, one evolves from a ``Higgs" 
description of the theory to a confining description, as the 
parameters of the model are changed, {\bf without} encountering 
a phase transition \cite{021}. 
A logical choice for $M_0$ is $M_0 \simeq \m$, as $\m$ is the 
only mass-scale in the problem.  Then the magnitude of 
$g^2 (\m ) /16\pi^2$ 
determines whether a ``Higgs" or confinement description is 
more appropriate. In our case, this corresponds to the 
evolution from $g^2(\mu )/16\pi^2 \ll 1$ to larger values of 
$g^2 (\mu )/16\pi^2$.  Thus Fig. 2 is appropriate to the Higgs description of the theory, since the Landau singularity 
obtained from running $g^2$ appears in the far infrared region.

When $g^2(\mu ) /16\pi^2$ small and $\mu^2 < 0$, the 
asymptotically free 
phase leads to a vector meson spectrum that is well described 
by perturbative estimates of masses, {\it i.e.}, N(N--2) 
massless vectors and 2N--1 massive vectors with $M^2_v \sim 
2 g^2 v^2$.  If on the other hand $g^2 (\mu^2 )/ 16\pi^2$ 
is large, then the vector mesons are strongly interacting 
at the characteristic mass-scale $\mu$, so that the 
confinement description would be more appropriate.  In 
analogy with the quark model, 
one then might wish to assign ``short-distance" masses to 
the vectors, 
and run them by means of a renormalization group.  However, 
this is well 
outside the scope of our calculation. In the asymptotically 
free phase, but with $\mu^2 > 0$, the gauge bosons should 
be regarded as massless, and confined.

The methods of this paper demonstrate that the two-phase 
structure is essential for understanding the physics.

\noindent{\bf C. Phase Transition in $N$?}

The question arises as to whether these results are stable
as $N$ is decreased.\footnote{We thank the referee for 
raising this question.}  We do not have the tools to answer 
this question in 
general, as our calculation is to all orders in $\l$,
but only leading orders in $1/N$ and $g^2$.  The complete
phase surface is inaccessible as it depends to all orders
on $\l$, $g^2$, and $1/N$.  We can provide a very limited
answer to this question by including the known $1/N$
corrections to the $\b$-functions to leading order in $g$.
Then instead of (2.5a) and (2.5b), we have 
\cite{013,014}

$$
16\pi^2 \b_{g^2} = - g^4 \left(\frac{44}{3} -
\frac{8}{3} \: \frac{N_f}{N} - \frac{2}{3N} \right)
+ {\cal O} (g^6)
\eqno{(3.10a)}
$$
$$
16\pi^2 \b_{1/\l} = 12 \left( 1 - \frac{1}{N^2} \right)
\: \frac{g^2}{\l} - 2 \left( 1 + \frac{4}{N} \right) \; .
\eqno{(3.10b)}
$$
It is straightforward to show that the phase-boundary
determined by (3.10) is
\renewcommand{\theequation}{3.\arabic{equation}}
\setcounter{equation}{10}
\be
0 < \l = 
\frac{ \left[ \frac{4}{3} \left( \frac{N_f}{N} -1 \right)
+ \frac{1}{3N} - \frac{6}{N^2} \right] g^2}
{\left( 1 + \frac{4}{N} \right)}
\ee
while asymptotic freedom for $g^2$ requires
\be
\left( \frac{11}{2} - \frac{1}{4N} \right) >
\frac{N_f}{N} \; .
\ee
Thus, if $N_f$ is fixed (and very large), then as $N$
is decreased (3.12) will eventually be violated, and
asymptotic freedom will be lost.  On the other hand,
if one keeps $ (\frac{N_f}{N} )$ fixed, so that (3.12) is
satisfied, then for $ (\frac{N_f}{N} ) > 2$, (3.11) can be
satisfied for $\l > 0$ for any $N$, so that the two-phase
structure of the model exhibited in the large $N$ limit
can be preserved.

\noindent{\bf 4. Gauge Invariance}

It is known for some time that the effective potential is 
not gauge invariant \cite{015,016}.  How does this impact 
our results?  We address this issue in this section.

\renewcommand{\theequation}{4.\arabic{equation}}
\setcounter{equation}{0}

Consider the effective \ul{action} 
$\G (\phi, \x, A_\m)$, and
\be
V(\bar{\phi}, \bar{\x}, \bar{A}_\m ) = - \int d^4x \, 
\G \left.
(\bar{\phi}, \bar{\x}, \bar{A}_\m )
{\rule{0mm}{4mm}}
\right/  \int d^4 x
\ee
when evaluated at the stationary points $\bar{\phi}, 
\bar{\x},
\bar{A}_\m$
defined by
\be
\frac{\d \G}{\d\phi} = \frac{\d \G}{\d\x} = 
\frac{\d \G}{\d A_\m} = 0 \; .
\ee
In this paper we have considered $V(\bar{\phi}, 
\bar{\x}, 0 )$, which is obtained from
\be
\left( \frac{\d \G}{\d\phi} \right)_{A{_\m}=0} = 
\left( \frac{\d \G}{\d\x} \right)_{A{_\m}=0} = 0
\ee
which need not be the stationary points of (4.2).  However 
it has been shown that by Fukuda and Kugo \cite{016} that 
there are a wide class of 
``\ul{good gauges}" where  $V(\bar{\phi}, \bar{\x}, 0 )$ 
correctly describes the stationary 
points of (4.2) by means of (4.3) with $\bar{A}_\m = 0$.  
The ``good gauges" include covariant gauges, Landau gauge, 
$R_\xi$ gauge, axial gauge \ldots .
[By contrast in a bad gauge, $\bar{A}_\m (x) \neq 0$ at the
stationary points, and $\bar{\phi} (x)$ has $x$ dependence to compensate
that of $\bar{A}_\m (x)$ and restore Lorentz invariance.] The 
Landau gauge employed in this paper is a good gauge, for which 
the stationary points are gauge invariant.

In more detail, it can be shown \cite{016} that the total
variation with respect to the gauge parameter satisfies a renormalization group type equation, where schematically
\begin{eqnarray}
\frac{D}{D\a} \: V(\phi , \a ) & = &
\left[ \frac{\pa}{\pa\a} - \g^{(\a)}_\l
\left( \l \frac{\pa}{\pa \l} \right) 
- \g^{(\a )}_\m 
\left( \m \frac{\pa}{\pa \m} \right) \right]
V(\phi , \a ) \nonumber \\[.05in]
& = & 
\left( \frac{\pa V}{\pa \phi_i} \right)
F_i (\phi, \a , A_\m ) \hspace{.4in} (i = 1\; \mbox{\rm to N})
\end{eqnarray}
where $F(\phi, \a, A_\m )$ is a functional of the fields and
\be
\g^{(\a )} = Z \: \frac{\pa}{\pa \a_0} \: \ln Z
\ee
for the two anomalous dimensions.  Thus
\be
\frac{D}{D\a} \: V(\phi , \a ) = 0 \hspace{.4in} \mbox{\rm at~~}
\phi = \bar{\phi} \; .
\ee
This means that the explicit gauge dependence cancels the 
implicit gauge
dependence of the parameters $(\m^2 / \l )$ and $\l$ \ul{at} 
the critical
point $\bar{\phi}$.  Therefore, the \ul{value} of the effective potential is gauge invariant at the critical points, so that 
one can select the 
critical point with the lowest value of the effective potential 
$V$ in a gauge invariant way  \cite{015}.  [Spontaneous 
symmetry breaking is a gauge invariant concept.]

Further,
\begin{eqnarray}
\frac{D}{D\a} \: \left[ \frac{\pa V(\phi ,\a)}{\pa \phi_j}
\right] & = &
\frac{\pa^2V}{\pa \phi_i \pa \phi_j} \; F_i (\phi , \a , A_\m )
\nonumber \\
[.05in]
& + & \frac{\pa V}{\pa \phi_i} \; \frac{\pa F_i}{\pa \phi_j} \;
(\phi , \a , A_\m )\\[.1in]
& = &
\left[ 2 \d_{ij} \left( \frac{\pa V}{\pa \phi^2} \right) \; + \;
4 \phi_i\phi_j \left( \frac{\pa^2 V}{(\pa \phi^2 )^2} \right)
\right]
F_i (\phi , \a , A_\m ) \nonumber \\[.05in]
& + & 
\left[ 2 \phi_i \left( \frac{\pa V}{\pa \phi^2} \right) \right]
\;
\frac{\pa F_i(\phi , \a , A_\m )}{\pa\phi_j} \;  \; .
\end{eqnarray}
At the phase-boundary, both
\be
\left( \frac{\pa V}{\pa \phi^2} \right)_{\bar{\phi}} = 0
\hspace{.5in} {\rm and}
\hspace{.5in}
\left( \frac{\pa^2 V}{(\pa \phi^2)^2} \right)_{\bar{\phi}} = 0
\ee
Hence at the phase boundary,
one also has at the stationary point of $V$,
\be
\frac{D}{D\a} \: \left[ \frac{\pa V}{\pa \phi_j}
\right]_{\bar{\phi}}
=0  \; .
\ee
This means that at the phase-boundary, not only is 
$V(\bar{\phi})$ gauge invariant, but 
$\left( \frac{\pa V}{\pa \phi} \right)_{\bar{\phi}}$ is as 
well. Thus, the vanishing of $dV/d\phi^2$ evaluated at 
$\bar{\phi}$ at the 
phase-boundary is a gauge invariant criterion, as expected 
for a zero-mass bound-state \cite{015}.

In general, the effective potential is \ul{not} gauge 
invariant \cite{015,016} so that the effective potential 
need not have the
specific behavior of Figs. 2 and 3 in other gauges.  However, 
the separation of the theory into asymptotically free and non-asymptotically free phases is a gauge invariant concept. 
Thus one expects the resolution 
of the difficulties of the ungauged vector model provided 
by the asymptotically free phase to be a physical feature 
of the model.

\newpage

\noindent{\bf Conclusions}

We have presented a calculation of the renormalization 
group improved effective potential for the gauged vector 
model coupled to $N_f$ massless fermions in the defining 
representation, computed to leading
order in 1/N, all orders in $\l$, and leading order in 
$g^2$.  It was shown that the theory has two phases.  
In the asymptotically free phase, the
effective potential behaves qualitatively like that of 
the tree-approximation, 
but with a Landau pole in the infrared region.  If $\l$ 
is too large, asymptotic freedom is destroyed, and the 
effective potential exhibits all 
the difficulties found previously for the
ungauged theory $(g^2 = 0)$ \cite{004}.

\newpage

\noindent{\bf Figure Captions}

\begin{itemize}
\item[\bf Fig. 1:] The $\l$ versus $g^2$ plane, and 
renormalization group 
flows.  Note the two phase structure.  The arrows point 
toward the ultraviolet.  The phase boundary is 
$\l = \frac{4}{3} (\frac{N_f}{N} - 1)
g^2$.  The graph is for $\frac{N_f}{N} = \frac{11}{4}$.
\item[\bf Fig. 2:] The real part of the effective potential 
vs. $\phi^2$ for $g^2/16\pi^2 = 0.01$, $\l /16\pi^2 
= 0.01$, 
\mbox{$\m^2/\l = -1$}, and $M = 1$, for $\frac{N_f}{N} 
= \frac{11}{4}$. In both Figs. 2 and 3, there are 
singularities (Landau poles) in $V$, near $\phi^2 = 1$,
$V=0$.  These features are not visible in the figures because
they occur in a range of $\phi^2$ which is
many orders of magnitude smaller than the scale displayed.
\item[\bf Fig. 3:] Same as Fig. 2, except $\l /16\pi^2 
=  0.20$.
\end{itemize}

\newpage
\pagestyle{empty}
\epsfbox{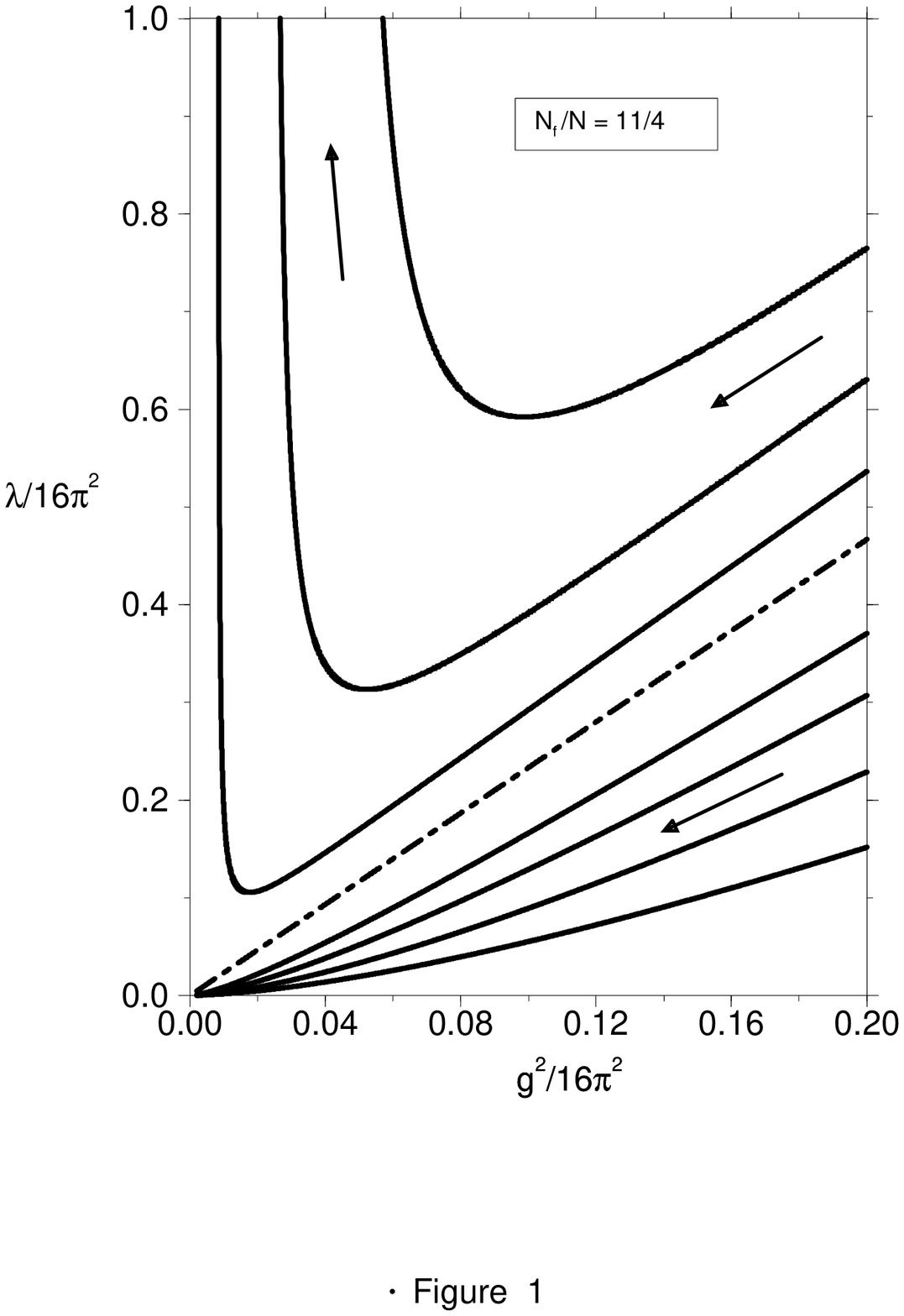}
\epsfbox{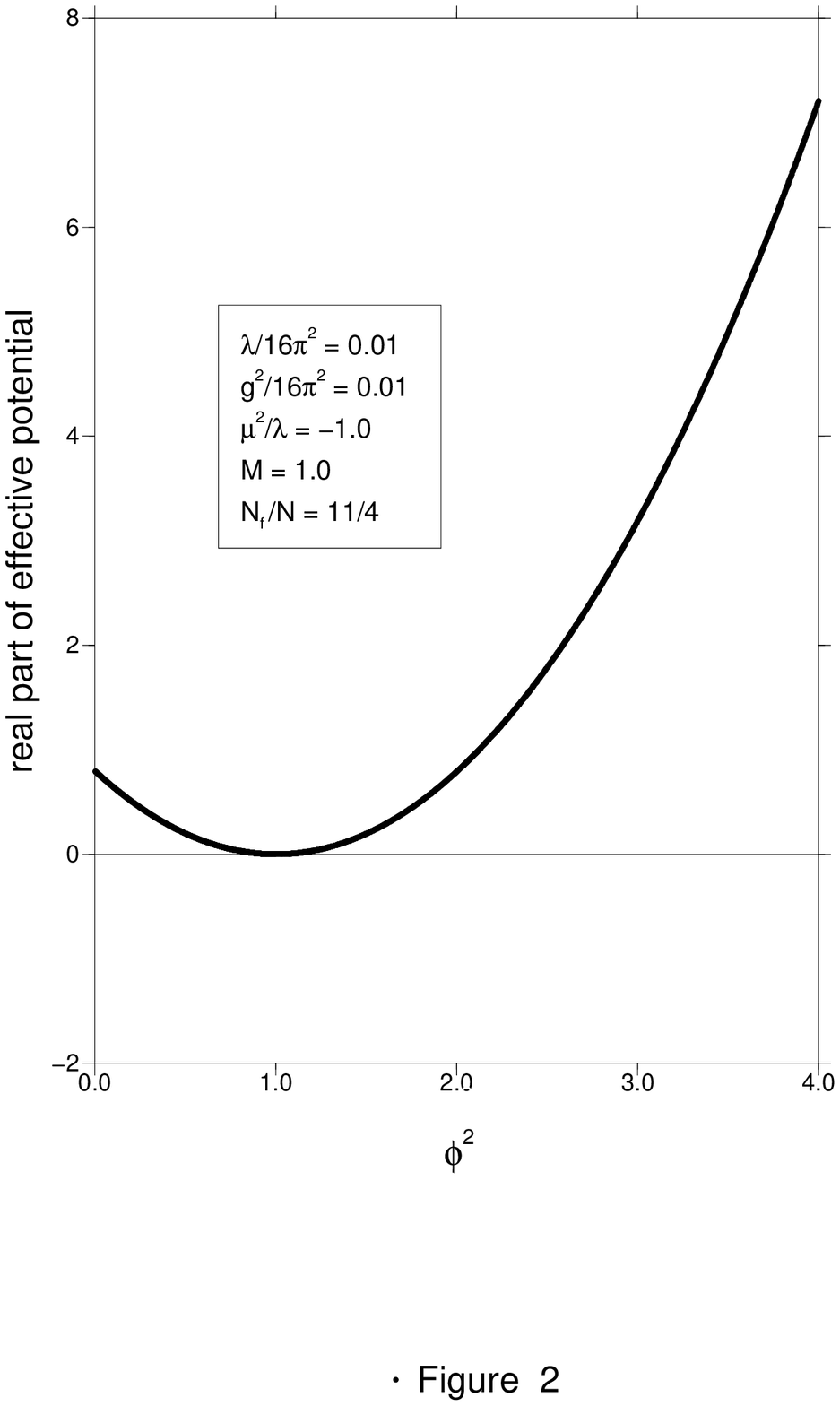}
\epsfbox{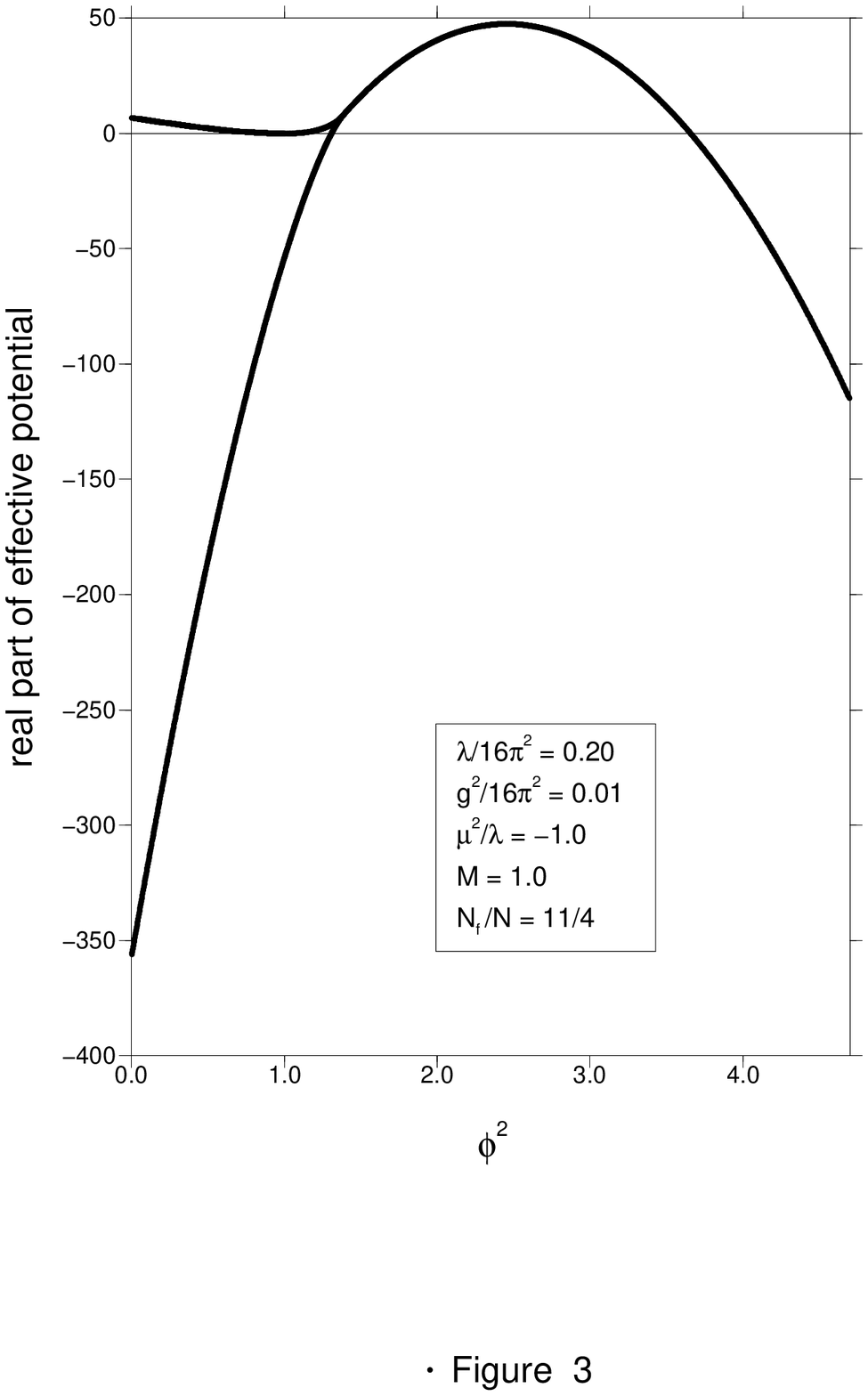}

\end{document}